# The Orientation of the Local Interstellar Magnetic Field

M. Opher[1*], E. C. Stone[2] & T. I. Gombosi[3]

The orientation of the local interstellar magnetic field introduces asymmetries in the heliosphere that affect the location of heliospheric radio emissions and the streaming direction of ions from the termination shock of the solar wind. We combine observations of radio emissions and energetic particle streaming with extensive 3D MHD computer simulations of magnetic field draping over the heliopause to show that the plane of the local interstellar field is ~ 60°-90° from the galactic plane. This suggests that the field orientation in the Local Interstellar Cloud differs from that of a larger scale interstellar magnetic field thought to parallel the galactic plane.

The heliosphere created by the supersonic solar wind is compressed by the motion of the Sun relative to the local interstellar medium, producing a comet-like shape with an extended tail. The solar wind abruptly slows, forming a termination shock as it approaches contact with the interstellar medium at the heliopause. Beyond the heliopause, the interstellar wind contains mainly hydrogen and helium, both as neutral atoms and as ions that carry the frozen-in interstellar magnetic field.

Recent Voyager observations of ions streaming from the termination shock (*1, 2*) have led to the suggestion that north/south and east/west asymmetries of the heliosphere are induced by the interstellar magnetic field (*3*). However, the inferred field direction from the model of (*3*) was parallel to the hydrogen deflection plane (HDP) rather than the galactic plane (GAL). Based on the polarization of light from nearby stars, (*4, 5*) suggested that the galactic magnetic field is parallel to the galactic plane. However, the direction of the galactic magnetic field is deduced from measurements over a much larger averaging volume (light years). A direction parallel to the HDP was suggested by (*6*) for the local interstellar field, based on the solar Lyman-$\alpha$ radiation that is resonantly backscattered by interstellar hydrogen atoms. The HDP is tilted from the ecliptic plane by 60° and differ from the GAL by 60°. We use Voyager 1 and 2 observations in conjunction with a magnetohydrodynamic model to discriminate between these two planes and constrain the orientation of the local interstellar magnetic field.

---

[1]. George Mason University, 4400 University Drive, Fairfax, VA 22030
[2]. California Institute of Technology, Pasadena, CA 91125
[3]. Center for Space Environment Modeling, University of Michigan, Ann Arbor, MI

[*]To whom correspondence should be addressed. Email: mopher@physics.gmu.edu

In the last 20 years, Voyager 1 (V1) and 2 (V2) have been detecting radio emissions in the outer heliosphere at frequencies from 2 to 3 kHz *(7-9)*. The radio emissions were detected each solar cycle: first in 1983-84 during the solar cycle 21 (*7*), second in 1992-94 during solar cycle 22 (*8*), and most recently in solar cycle 23 (*9*). The currently accepted scenario is that the radio emissions are generated when a strong interplanetary shock produced by a period of intense solar activity reaches the vicinity of the heliopause and move into the interstellar plasma beyond (*9, 10*). Radio direction-finding measurements from V1 and V2 have been used to determine the positions near the heliopause at which the radio emission are generated (*11*) (Fig. 1). The sources lie along a line that passes near the nose of the heliosphere that roughly parallels the galactic plane. The GAL plane is 120° from the ecliptic plane (see SOM text 1). Based on the fact that the galactic magnetic field is oriented nearly parallel to the galactic plane, (*11*) suggested the local interstellar magnetic field (in the local neighborhood of the sun) was also parallel to the galactic plane.

However, (*12*) recently pointed out that at the earth's bow shock and interplanetary shocks, the radio emission occurs where the magnetic field lines are tangential to the shock surface and suggested that heliospheric radio emissions occur where the local interstellar magnetic field is tangential to the surface of the shock that excites the plasma (or **B**•**n**=0, where **B** is the magnetic field and **n** is the shock normal). They conclude that the condition **B**•**n**=0 combined with the observed source location by Voyager spacecrafts implies that the local interstellar magnetic field is perpendicular to the galactic plane. This direction differs from the earlier the earlier suggestion (*9*) and is within 16° of the HDP plane.

The interstellar magnetic field is frozen into interstellar plasma that is deflected around the heliopause, causing the field to drape over the heliopause. As a result, the region where **B**•**n**=0 will depend on the shape of the heliopause, which is distorted by pressure of the local interstellar magnetic field. For intensities around a few microgauss, the ambient interstellar magnetic pressure is comparable to the gas pressure, with the magnetic pressure increasing in those regions where the interstellar flow decreases as it approaches the heliopause. We investigated how the proposed location of the radio sources (where **B**•**n**=0 on the surface of the heliopause) varies with the orientation and strength of the local interstellar magnetic field.

We considered several directions of interstellar magnetic field: the hydrogen deflection plane; the galactic plane; and the plane perpendicular to the galactic plane (PPG) with different inclination angles $\alpha$, where $\alpha$ is the angle between the interstellar magnetic field and interstellar wind velocity. In the model coordinate system, with $\beta$ being the angle between the interstellar magnetic field and the solar equator, HDP correspond to $\beta=60°$, GAL to $\beta=120°$, and PPG to $\beta=44°$ (see SOM text 1). Assuming a spherical interplanetary shock, the tangential field condition for the radio sources translates to $B_r=0$ with $B_r$ being the radial component of the interstellar magnetic field. For each modeled direction of the interstellar magnetic field, we compared the expected location of the radio sources ($B_r=0$ at the heliopause) with the observed location of the radio sources detected by V1 and 2.

The model used here is the same as used by (*3*) (see SOM text 2). The interstellar magnetic field ($B_{ISM}$) magnitude is taken to be $B_{ISM}=1.8\mu G$ (with the y component of $B_{ISM}$, $B_{ISM,y}<0$). The coordinate system has the interstellar velocity direction in the +x direction and the z-axis as the solar rotation axis of the sun, with y completing the right handed coordinate system. In this coordinate system, V1 is at

29.1° in latitude and 213.4° in longitude and V2 is at -31.2° and 178.4° in longitude, which ignores the 7.25° tilt of the solar equator with respect to the ecliptic plane.

Fig. 1 indicates that the heliopause is strongly influenced by the interstellar magnetic field direction; the heliopause is asymmetric both north/south and east/west and has a plane of symmetry approximately parallel to the plane of the local interstellar magnetic field. As a result, the heliopause surfaces for HDP and GAL field orientations are almost mirror images of each other.

With $B_{ISM}$ parallel to the galactic plane (GAL, with $\alpha$=45°, Fig1d), the region where $B_r$=0 is almost perpendicular to the galactic plane, which is inconsistent with the radio observations. An interstellar magnetic field perpendicular to the galactic plane (PPG plane, with $\alpha$=30°, Fig1c) produces the best agreement with the radio observations by Voyager, as suggested by (*12*). The HDP orientation differs from PPG by only 16° and is also in general agreement as suggested by the similarity of the regions with $B_r$=0 in the upper left (HDP) and lower left (PPG) panels. The offset of ~15° between the observations and the region with $B_r$=0 for the model in best agreement (Fig. 1c) indicates that the accuracy of the model is not adequate to distinguish between the PPG and HDP field orientations.

We also investigated the effect of changing the interstellar wind direction to 5° above the ecliptic plane (in the solar ecliptic coordinate system, the interstellar wind direction is 255° (longitude) and 5° (latitude)), and changed the intensity of $B_{ISM}$ from 1.8μG to 2.5μG. For both cases, the change in the predicted of radio sources location was minor. As $\alpha$ increases from 15° to 60°, the $B_r$=0 band moves counterclockwise, with the best agreement for $\alpha$=30°-45° (see SOM text3).

The second set of observational data that we used to constrain the orientation of the local interstellar magnetic field was the streaming ions from the termination shock. V1 crossed the termination shock at 94 AU in December 2004 and is now beyond 100 AU in the heliosheath (*1-2, 13*). V2 is already detecting signs of the upcoming shock (*2, 14*) and is expected to cross the termination shock in the next 1-2 years. In mid 2002, V1 began observing enhanced intensities of ions streaming from the shock (*15, 16*). The beams of energetic termination shock particles (TSPs) were streaming outward along the solar spiral magnetic field. The strong upstream TSP beams were observed much of the time until V1 crossed the shock at 94 AU. The streaming along the magnetic field upstream of the shock source was expected to be inward along the spiral field if the termination shock is spherical. With a non-spherical shock, Voyager 1 could be connected to the termination shock along magnetic field lines that crossed the termination shock and crossed back to the supersonic solar wind. However, the observed flow was outward along the field, requiring a shock source located inward along the spiral field several AU closer to the Sun than V1. This led to the suggestion that the upstream beaming resulted from a blunt (*17*) or asymmetric shock (*3*). The asymmetric shock could result from an interstellar magnetic field inclined in a particular direction (*18, 19*). In a recent work (*3*), we showed that an interstellar magnetic field in the HDP could distort the termination shock in a direction that explains the TSPs outward streaming at V1.

Fig 2. shows that for $B_{ISM}$ parallel to the HDP, the longitude of the point MD at which the shock has the minimum radial distance to the Sun is greater than the longitude of V1, so the TSPs will stream outward along the spiral field. In the southern hemisphere the longitude of V2 is greater than that of the MD point of the shock, so the TSPs will stream inward toward V2, as is observed. However, for $B_{ISM}$ parallel to the galactic plane GAL, the MD in the northern hemisphere shifts to a smaller longitude than V1, so that the TSPs would stream inward toward V1, opposite to what is observed.

In this calculation we did not include the neutral hydrogen atoms that interact with the ionized component by charge exchange. Although the inclusion of the neutral atoms will tend to symmetrize the solution and quantitatively affect the degree of asymmetry, the general character of the asymmetry is expected to remain the same, with the plane of symmetry of the distorted heliopause determined by the plane of the local interstellar magnetic field (*20, 21*). Thus, it would be expected that different orientations of the local interstellar magnetic field would result in the same qualitative differences in the predicted radio source locations and streaming directions of upstream ions as described here. Based on those differences, and assuming that the source of radio emission is the region where the field draped on the heliopause is perpendicular to the radial direction, we find from Voyager observations that the plane of the local interstellar magnetic field is not parallel to the galactic plane, but 60° to 90° from that plane (rotated clockwise from a view from the Sun). This suggests that the field orientation in the Local Interstellar Cloud differs from that of a larger scale interstellar magnetic field thought to parallel the galactic plane.

The authors would like to thank the staff at NASA Ames Research Center for the use of the Columbia supercomputer.

# The Orientation of the Local Interstellar Magnetic Field

M. Opher[1], E. C. Stone[2] & T. I. Gombosi[3]

The orientation of the local interstellar magnetic field introduces asymmetries in the heliosphere that affect the location of radio emission and the streaming direction of ions from the termination shock of the solar wind.   Comparing a magnetohydrodynamic model with Voyager observations, we find that the plane of the local interstellar field is ~ 60°-90° from the galactic plane.  This suggests that the field orientation in the Local Interstellar Cloud differs from that of a larger scale interstellar magnetic field thought to parallel the galactic plane.

Recent Voyager observations have led to the suggestion of north/south and east/west asymmetries of the heliosphere are induced by the interstellar magnetic field (1). However, the inferred field direction was parallel to the hydrogen deflection plane (HDP) rather than the galactic plane (GAL) as suggested by Frisch (1990, 1996) *(2,3)* based on the polarization of light from nearby stars. The HDP was suggested by *Lallement et al. (2005) (4),* based on the solar Lyman-$\alpha$ radiation that is resonantly backscattered by interstellar hydrogen atoms. The HDP is tilted from the galactic plane by 60°. Using Voyager 1 and 2 observations in conjunction with a magnetohydrodynamic model, we are able to discriminate between these two planes and predict the direction of the local interstellar magnetic field.

In the last 20 years, Voyager 1 (V1) and 2 (V2) have been detecting radio emissions in the outer heliosphere at frequencies from 2 to 3 kHz *(5-7)*. The radio emissions were detected during each solar cycle. Gurnett et al. (2006) (*8*) suggested a new way to constrain the local interstellar magnetic field based on the radio emission detected by Voyager 1 and 2. The currently accepted scenario is that the radio emissions are generated when a strong interplanetary shock produced by a period of intense solar activity interacts with the heliopause (*6,9*). Radio direction-finding measurements from V1 and V2 have been used to determine the positions near the heliopause at which the radio emission are generated (*10*) and are shown in bottom panels of Fig. 1. The sources lie along a line that passes near the nose of the heliosphere at an angle roughly parallel to the galactic plane. Based on the fact that the galactic magnetic field is oriented nearly parallel to the galactic plane, Kurth and Gurnett (*10*) suggested the local interstellar magnetic field was also parallel to the galactic plane.

However, Gurnett et al. (2006) (*8*) recently pointed out that at the earth's bow shock and interplanetary shocks, the radio emission occurs where the magnetic field lines are tangential to the shock surface and suggested that heliospheric radio emissions occur where the local interstellar magnetic field is tangential to the surface of the shock that excites the plasma (or **B**•**n**=0, where **B** is the magnetic field and **n** is the shock normal). They conclude that the condition **B**•**n**=0 combined with the observed source location implies that the local interstellar magnetic field is perpendicular to the galactic plane.

There are several studies investigating the concept that the heliospheric 2-3kHz radio emission is generated just beyond the heliopause by the same two-step beam-plasma mechanism as in interplanetary shocks (e.g., (*11,12*)). Because the interstellar field is draped over the heliopause, the region where **B**•**n**=0 will depend on the shape of the heliopause, which is distorted by the local interstellar magnetic field. For intensities around a few microgauss, the ambient interstellar magnetic pressure is comparable to the gas pressure, with the pressure increasing in those regions where the interstellar flow decreases as it approaches the heliopause. We investigated how the proposed location of the radio sources (**B**•**n**=0) varies with the orientation and strength of the local interstellar magnetic field.

We considered several directions of interstellar magnetic field: the hydrogen deflection plane ; the galactic plane; and the plane perpendicular to the galactic plane (PPG) with different inclination angles $\alpha$, where $\alpha$ is the angle between the interstellar magnetic field and interstellar wind velocity. For each modeled direction of the interstellar magnetic field, we compared the expected location of the radio sources taken at the heliopause with the observed location of the radio sources detected by V1 and 2. Assuming a spherical interplanetary shock, the tangential field condition for the radio sources translates to $B_r$=0 with $B_r$ being the radial component of the interstellar magnetic field.

The model used here is the same as used by Opher et al. (2006) (*1*). It is based on the BATS-R-US code, a three-dimensional MHD parallel, adaptive grid code developed at the University of Michigan (*25*) and adapted by Opher et al. (2003, 2004) (*13,14*) for the outer heliosphere. We used the same grid and parameters as used by Opher et al. (2006). The interstellar magnetic field ($B_{ISM}$) magnitude is taken to be $B_{ISM}$ =1.8nT (with the y component of $B_{ISM}$, $B_{ISM,y}$<0). The coordinate system has the interstellar velocity direction in the +x direction and the z-axis as the solar rotation axis of the sun, with y completing the right handed coordinate system. In this coordinate system, V1 is at 29.1° in latitude and 213.4° in longitude and V2 is at -31.2° and 178.4° in longitude, which ignores the 7.25° tilt of the solar equator with respect to the ecliptic plane.

Fig. 1 shows the predicted location of the radio sources as a function of interstellar magnetic field direction. The upper two panels show the heliopause iso-surface with contours of the radial component of the interstellar magnetic field for $B_{ISM}$ in the HDP and GAL plane (both with $\alpha$=45°). The green band indicates where $B_r=0$, the suggested source location of the heliospheric radio emissions. The figure indicates that the heliopause is strongly influenced by the interstellar magnetic field direction; the heliopause is asymmetric both north/south and east/west and has a plane of symmetry approximately parallel to the plane of the local interstellar magnetic field. As a result, the heliopause surfaces for HDP and GAL field orientations are almost mirror images of each other.

The lower two panels show the $B_r$ isocontours in ecliptic longitude and latitude coordinates for comparison with the source locations of the radio emissions, with $B_{ISM}$ in the: PPG (with $\alpha=30°$) and GAL (with $\alpha=45°$). For $B_{ISM}$ parallel to the galactic plane (GAL, with $\alpha=45°$, bottom right panel), the region where $B_r=0$ is almost perpendicular to the galactic plane, inconsistent with the radio observations. An interstellar magnetic field perpendicular to the galactic plane (PPG plane, with $\alpha=30°$, lower left panel) produces the best agreement with the observations, as suggested by Gurnett et al (2006). The HDP orientation differs from PPG by only 16° and is also in general agreement as suggested by the similarity of the regions with $B_r=0$ in the upper left (HDP) and lower left (PPG) panels. The offset of ~15° between the observations and the region with $B_r=0$ for the model in best agreement (Figure 1c) indicates that the accuracy of the model is not adequate to distinguish between the PPG and HDP field orientations.

We also investigated the effect of including the fact that the interstellar wind direction is 5° latitude above the ecliptic plane (in the solar ecliptic coordinate system, the interstellar wind direction is 255° (longitude) and 5° (latitude)), and varied the intensity of $B_{ISM}$ (from 1.8nT to 2.5nT). For both variations, the change in the predicted of radio sources location was minor. As $\alpha$ increases from 15° to 60°, the $B_r=0$ band moves counterclockwise and the best agreement is for $\alpha=30°-45°$.

The second set of observational data that we used to constrain the orientation of the local interstellar magnetic field was the streaming ions from the termination shock. V1 crossed the termination shock at 94 AU in December 2004 and is now beyond 100 AU in the heliosheath(*16-19*). V2 is already detecting signs of the upcoming shock (*18,19*) and is expected to cross the termination shock in the next 1-2 years. In mid 2002, V1 began observing enhanced intensities of ions streaming from the shock (*20-21*). The beams of energetic termination shock particles (TSPs) were streaming outward along the spiral magnetic field. The strong upstream TSP beams were observed much of the time until V1 crossed the shock at 94 AU. The streaming along the magnetic field upstream of the shock source was expected to be inward along the spiral field if the termination shock is spherical. However, the observed flow was outward along the field, requiring a shock source located inward along the spiral field several AU closer to the Sun than V1. This led to the suggestion that the upstream beaming resulted from a blunt (*22*) or non-spherical shock (*18*)). The nonspherical shock could result from an interstellar magnetic field inclined in a particular direction (*23-24*). In a recent work (*1*), we showed that an interstellar magnetic field in the HDP could distort the termination shock in a direction that explains the TSPs outward streaming.

In Fig. 2, we show the direction of streaming of the TSPs expected for orientations of $B_{ISM}$ : parallel to the hydrogen deflection plane, HDP and to galactic plane, GAL. V1 and V2 are first connected along the spiral magnetic field lines to the termination shock at the minimum distance of the shock to the Sun (labeled MD). Because the magnetic field is carried radially outward by the solar wind, the field forms a spiral on a conical surface. In the upper two rows the green line indicates the location of the non-spherical termination shock and several spiral magnetic field lines on the cone intersecting the latitude of Voyager 1. Similar plots are shown for the conical surface intersecting the latitude of Voyager 2 in the southern hemisphere, and the predicted streaming directions of the TSPs are summarized in solar ecliptic coordinates in the bottom two panels.

For $B_{ISM}$ parallel to the HDP, the longitude of the point MD at which the shock has the minimum radial distance is greater than the longitude of V1, so the TSPs will stream outward along the spiral field. In the southern hemisphere the longitude of V2 is greater than that of the MD point of the shock, so the TSPs will stream inward toward V2, as is observed.  However, for $B_{ISM}$ parallel to the galactic plane GAL, the MD in the northern hemisphere shifts to a smaller longitude than V1, so that the TSPs would stream inward toward V1, opposite to what is observed.

   In this calculation we did not include the neutral hydrogen atoms that interact with the ionized component by charge exchange.  Although the inclusion of the neutral atoms will tend to symmetrize the solution and quantitatively affect the degree of asymmetry, the general character of the asymmetry is expected to remain the same, with the plane of symmetry of the distorted heliopause determined by the plane of the local interstellar magnetic field (*e.g., Figure 4c in Pogorelov et al. (25)* for a model with neutrals).   Thus, it would be expected that different orientations of the local interstellar magnetic field would result in the same qualitative differences in the predicted radio source locations and streaming directions of upstream ions as described here.  Based on those differences, and assuming that the source of radio emission is the region where the field draped on the heliopause is perpendicular to the radial direction, we find from Voyager observations that the plane of the local interstellar magnetic field is not parallel to the galactic plane, but 60° to 90° from that plane.  This suggests that the field orientation in the Local Interstellar Cloud differs from that of a larger scale interstellar magnetic field thought to parallel the galactic plane.

[1]. George Mason University, 4400 University Drive, Fairfax, VA 22030, Email: mopher@physics.gmu.edu
[2]. California Institute of Technology, Pasadena, CA 91125
[3]. Center for Space Environment Modeling, University of Michigan, Ann Arbor, MI

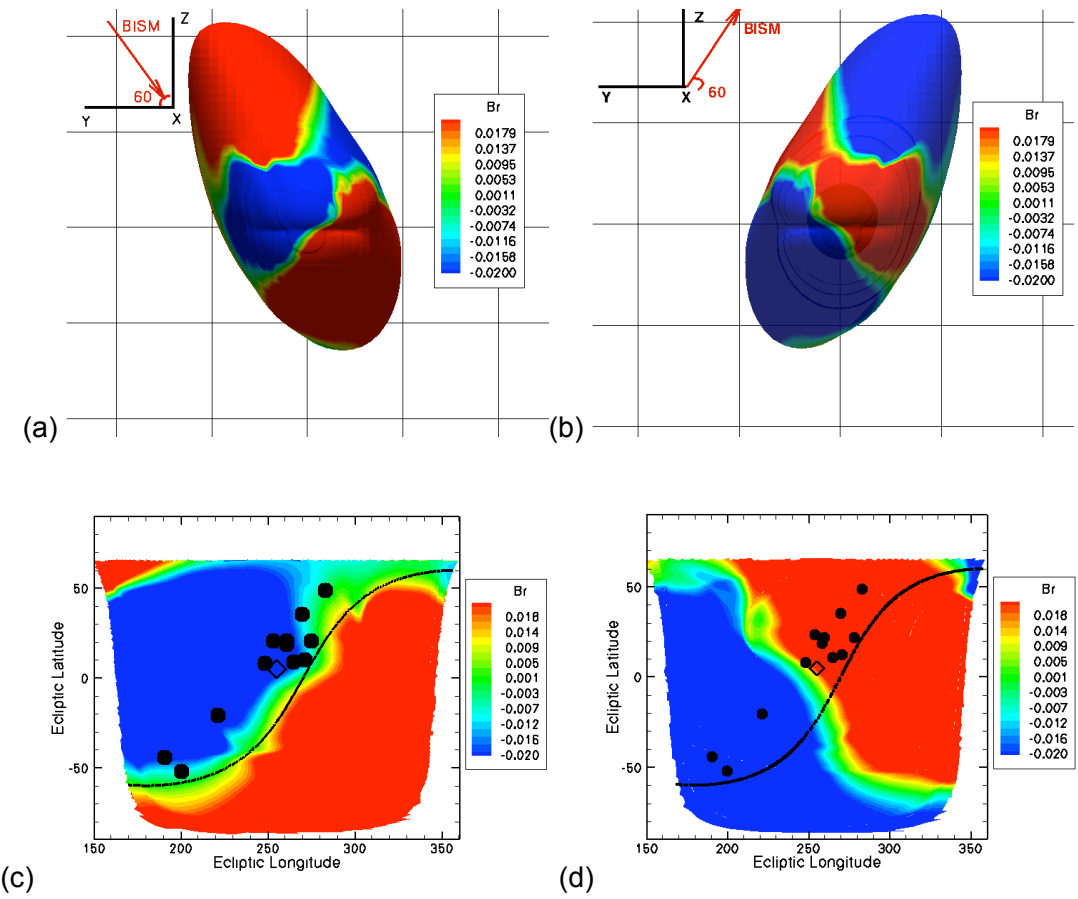

Fig 1.

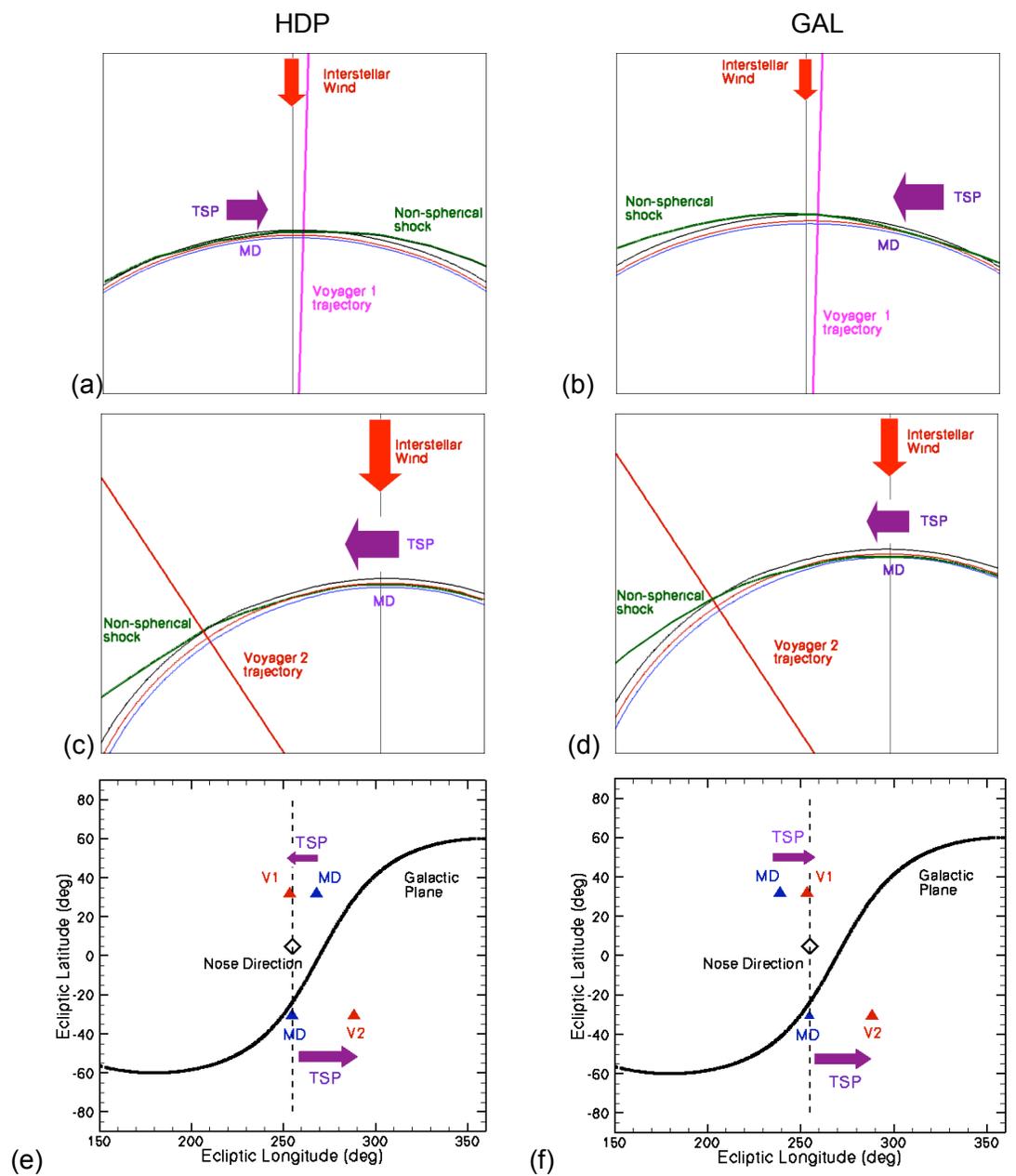

Fig 2.

**Fig 1.** The radio source location as a function of the interstellar magnetic field ($B_{ISM}$) direction for $B_{ISM}$, in the a) HDP plane and b) GAL plane (with $\alpha=45°$). Panels (a) and (b) shows the 3D iso-surface of the heliopause as viewed from upwind from the interstellar wind point of view. The iso-surface of the heliopause is shown with contours of the radial component of the interstellar magnetic field, $B_r$. The green band is the location of the radio sources (at $B_r=0$). The red arrows in panels (a) and (b) show the direction of $B_{ISM}$. Panels (c) and (d) show same as upper panels but converted to ecliptic coordinates for $B_{ISM}$ in the a) plane perpendicular to the galactic plane (with $\alpha=30°$) and b) GAL plane (with $\alpha=45°$). The direction of the nose of the heliosphere (diamond) and the galactic plane (black lines) are indicated for reference. The radio sources detected by V1 and V2 are shown (black points). Note that the color invert from the left image to the right image. This is because the interstellar magnetic direction was inverted from case (a) to (b) (see red arrows).

**Fig. 2.** Streaming of termination shock particles (TSPs) (magenta arrows) from the minimum distance of the termination shock to the Sun (MD) back to Voyager 1 (V1) and 2 (V2), for the interstellar magnetic field in the a) hydrogen deflection plane (with $\alpha=45°$) and b) galactic plane (with $\alpha=45°$). V1 and V2 are first connected along the spiral magnetic field lines to the termination shock at the minimum distance of the shock to the Sun (labeled MD). Because the magnetic field is carried radially outward by the solar wind, the field forms a spiral on a conical surface. In the (a)-(d) panels the green line indicates the non-spherical termination shock. The (a) and (b) panels shows the solar magnetic field lines that intersect Voyager 1; the field line intersecting the shock where V1 crosses the shock is labeled 0 AU (black) with red and blue indicating, respectively, magnetic field lines 2.0AU and 3.0AU upwind from the 0AU line. The magenta arrow indicates the streaming direction of the termination shock particles from the shock along the field line to V1. Panels (c) and (d) shows similar plot for V2, showing field lines 3.0 and 5.0AU upwind of the 0AU line. Note that in both views the solar magnetic field spirals clockwise with increasing distance outward. The panels (e) and (f) summarize the streaming of TSPs from MD back to V1 and V2. The nose direction (diamond) as well as the galactic plane.

**Fig 1.** The radio source location as a function of the interstellar magnetic field ($B_{ISM}$) direction for $B_{ISM}$, in the a) HDP plane and b) GAL plane (with $\alpha=45°$). Panels (a) and (b) shows the 3D iso-surface of the heliopause as viewed from upwind from the interstellar wind point of view. The iso-surface of the heliopause is shown with contours of the radial component of the interstellar magnetic field, $B_r$. The green band is the location of the radio sources (at $B_r=0$). The red arrows in panels (a) and (b) show the direction of $B_{ISM}$. Panels (c) and (d) show same as upper panels but converted to ecliptic coordinates for $B_{ISM}$ in the a) plane perpendicular to the galactic plane (with $\alpha=30°$) and b) GAL plane (with $\alpha=45°$). The direction of the nose of the heliosphere (diamond) and the galactic plane (black lines) are indicated for reference. The radio sources detected by V1 and V2 are shown (black points). Note that the color invert from the left image to the right image. This is because the interstellar magnetic direction was inverted from case (a) to (b) (see red arrows).

**Fig. 2.** Streaming of termination shock particles (TSPs) (magenta arrows) from the minimum distance of the termination shock to the Sun (MD) back to Voyager 1 (V1) and 2 (V2), for the interstellar magnetic field in the a) hydrogen deflection plane (with $\alpha=45°$) and b) galactic plane (with $\alpha=45°$). V1 and V2 are first connected along the spiral magnetic field lines to the termination shock at the minimum distance of the shock to the Sun (labeled MD). Because the magnetic field is carried radially outward by the solar wind, the field forms a spiral on a conical surface. In the (a)-(d) panels the green line indicates the non-spherical termination shock. The (a) and (b) panels shows the solar magnetic field lines that intersect Voyager 1; the field line intersecting the shock where V1 crosses the shock is labeled 0 AU (black) with red and blue indicating, respectively, magnetic field lines 2.0AU and 3.0AU upwind from the 0AU line. The magenta arrow indicates the streaming direction of the termination shock particles from the shock along the field line to V1. Panels (c) and (d) shows similar plot for V2, showing field lines 3.0 and 5.0AU upwind of the 0AU line. Note that in both views the solar magnetic field spirals clockwise with increasing distance outward. The panels (e) and (f) summarize the streaming of TSPs from MD back to V1 and V2. The nose direction (diamond) as well as the galactic plane.